\RequirePackage{fix-cm}
\documentclass[smallextended]{svjour3}
\usepackage{soul}
\usepackage{color}
\usepackage{amsfonts}
\usepackage{bm}%
\usepackage[colorlinks=true,linkcolor=blue,filecolor=blue,menucolor=yellow,urlcolor=blue,
citecolor=blue,anchorcolor=blue]{hyperref}
\usepackage{graphicx}% Include figure files
\usepackage{dcolumn}% Align table columns on decimal point
\usepackage{rotating}
\journalname{Journal of Low Temperature Physics}

\begin{document}

\title{Dislocation Mobility and Anomalous Shear Modulus Effect
in $^4$He Crystals} %otvam

\author{Abdul N. Malmi-Kakkada \and  Oriol T. Valls \and
Chandan Dasgupta} %abdiop0 adjusted according to iop
%\email{malmikakkada@physics.umn.edu}
\institute{Abdul N. Malmi-Kakkada \and Oriol T. Valls \at 
School of Physics and Astronomy, University of Minnesota, 
Minneapolis, Minnesota 55455 \\
\email{otvalls@umn.edu} \\
\emph{Present address:} of Abdul N. Malmi-Kakkada \at
 Dept. of Chemistry, University of Texas at Austin, \\
 105 E 24th St, Austin, TX 78712 \\
 \and Chandan Dasgupta \at
Centre for Condensed Matter Theory, Department of Physics,
Indian Institute  of Science, Bangalore 560012, India}

\date{\today}

\maketitle

\begin{abstract} 
We calculate the dislocation glide mobility in solid
$^4$He within a model that assumes the existence
of a superfluid field associated with dislocation lines. 
Prompted by the results of this  mobility calculation, 
we study within this model the role that such a
superfluid field  may play in the motion of the dislocation
line when a stress is applied to the crystal. 
To do this, we relate  the 
damping of dislocation motion, calculated 
in the presence of the assumed superfluid field, to
the shear 
modulus of the
crystal. As the temperature increases, 
we find that a sharp drop in the shear modulus will occur  at the 
temperature where the superfluid field disappears. 
We compare the drop in shear modulus of the crystal
arising from the temperature dependence of the damping contribution due
to the 
superfluid field,
to the experimental observation of the same phenomena in solid $^4$He
and find quantitative agreement. Our results indicate that
such a superfluid field  %sull4 last sentence entirely rewritten
plays an important role in dislocation pinning %sull5
in a clean solid $^4$He  at low
temperatures and in this regime may provide an alternative source for the
unusual elastic phenomena observed in solid $^4$He.
\end{abstract}

\section{Introduction}
% Need to account for the competing theories - Fefferman et. al. and Zhou et.al. 
Solid $^4$He is the archetype of a quantum crystal.
%due to the
%manifestation of quantum effects.
 Quantum
effects in 
solid $^4$He were pointed out as early as 1960s~\cite{hether,Andreev},
and extensive work on this and many other aspects
of its properties~\cite{solidreview}
has been subsequently performed.
Among the quantum mechanical effects 
it exhibits are those  associated with crystalline defects. %otvam
%characterized by large zero point vibration of atoms. 
More recently, the observation of a marked 
period drop with temperature in torsional oscillator 
experiments~\cite{kim,reppy} in solid $^4$He, originally interpreted
as evidence for a ``supersolid'' state, renewed both theoretical 
and experimental  
interest on topics related to quantum crystals. Subsequently, solid $^4$He
was also shown to undergo an anomalous softening of
the shear modulus~\cite{day,day1}. This drop in the shear modulus 
was observed at the same temperature range 
as the drop in period seen in 
torsional oscillator experiments. 
%Although supersolidity in solid $^4$He is highly controversial, 
These results suggested that the anomalous shear modulus effect, rather
than the change in the inertial mass dragged by the oscillator, was 
responsible for the observed drop in torsional oscillator 
period~\cite{kim1,iwasa}.  %otvam
Nevertheless, the discovery of 
this anomalous shear modulus softening has led to 
new and important questions being posed on the elastic
properties of solid $^4$He and in particular on the role  
of dislocation lines in this material. Given that the 
mechanical properties of crystals are largely dictated by 
dislocation lines, the observed anomalous shear modulus behavior can provide 
fundamental insights into the elastic properties of quantum crystals. 
%with superfluid field, impurities phonons etc. 
That is, some of the features associated with the 
behavior of the dislocation lines in solid
 $^4$He crystal may be due to quantum crystal effects and therefore not 
be ordinarily observed in classical crystals. 

Indeed, the reasons for the %sull3
anomalous shear modulus 
behavior in solid $^4$He are still much debated. %sull4 grammar, added still
A frequently  held position is simply that
dislocation lines are pinned by $^3$He impurities at low temperatures but 
become mobile (able to glide) at higher temperatures when impurities are no 
longer able 
to pin the dislocation network~\cite{day}. 
The experimentally observed dependence of the elastic anomaly on $^3$He %cdsull importance of impurities
concentration suggests that impurities play an important role in
this phenomenon.
Other 
proposals~\cite{iwasa,iwasa1,fefferman} model dislocation 
lines as 
vibrating strings unable to execute free glide 
motion. 
String-like bowing of such dislocation lines  in response to stress is 
taken into account 
in explaining the shear modulus behavior. 
Yet another proposal~\cite{zhou,zhou1}
attempts to model the shear modulus behavior by taking into account 
the interactions between
dislocation lines as well as the Peierls barrier contribution to the damping 
of the dislocation motion. 

The role that quantum
phenomena may play in the anomalous elastic
behavior is disputed. The effect that a putative quantum field
associated with   dislocation lines could play in the 
anomalous shear 
modulus behavior has been recently explored in the literature~\cite{kuklov}. 
Quantum Monte Carlo calculations~\cite{troyer,troyer1} have shown that 
superfluidity can occur in dislocation cores. 
A recent experimental study raised the possibility that 
dissipative mass flux in solid $^4$He could be associated with superfluid
cores of edge dislocations~\cite{vekhov}. Thus, the
question of the presence of 
superfluid field in solid $^4$He has become controversial, as seen 
on the one hand from 
the several models mentioned above, 
that seek to explain the anomalous shear modulus effect without 
invoking the presence of a superfluid field, and on the other  
by numerous other studies~\cite{kuklov,aleinikava,soyler} 
which take into account, or discuss 
the possibility of~\cite{jltp1,jltp2,newk,newchoi,newchoi1} 
the presence of a superfluid field.  %cdsull added reference
Therefore, it seems to us
highly pertinent to investigate the effect of 
an assumed superfluid field on dislocation
motion within crystalline $^4$He. 

Given this context of
exploring the role that superfluidity may play in solid $^4$He, 
we begin by examining the possible effects of an assumed superfluid field 
on its elastic properties 
via a novel calculation of
the  dislocation mobility in a quantum
crystal. Assuming a minimal coupling~\cite{toner}
between the dislocation line strain 
and the superfluid field it is in principle possible for 
the superfluid field either to make 
it easier for the dislocation line to move or 
to  contribute to the 
pinning of the dislocation line, making it harder for it to move.
A prior  study of  
the interplay of dislocation lines and superfluid field showed that 
a quenched dislocation line enhances superfluidity near it while  
moving dislocation lines can suppress superfluidity in its 
vicinity~\cite{malmi}.
The results of our calculation 
of  %the  contribution to  %sull3 edit here
dislocation motion damping arising from 
an assumed superfluid field associated with the dislocation line, %sull3 typo
lead us to investigate the possibility that at low temperatures and in 
low impurity crystals,
such damping may play
an important role in pinning the dislocation motion and 
therefore affect the shear modulus behavior. 
At low temperatures, we would expect quantum 
effects other than thermal phonon scattering and impurity pinning
to be more important in terms of damping of dislocation line motion and the 
associated anomalous shear modulus behavior.  
Experimental studies of ultra pure solid $^4$He samples with negligible 
concentration of $^3$He impurities also
exhibit anomalous shear modulus behavior~\cite{souris,pantalei}. 
For $^4$He crystals 
characterized by distances between $^3$He
impurity atoms larger than the cell size in which 
$^4$He is contained, anomalous shear softening is
reported~\cite{rojas}.
High quality $^4$He crystals presumably with low 
impurity concentration, 
also exhibit anomalous shear modulus behavior~\cite{haziot1}.  
These experimental observations inevitably point out, 
as noted in Ref.~\cite{pantalei}, that 
dislocation pinning by impurities may not be the only mechanism responsible
for anomalous shear modulus behavior.

In this  paper, then, we 
explore the consequences, for the elastic properties
of a quantum crystal, of assuming the existence of a
superfluid field associated
with the dislocation lines and 
examine its relation to the sharp decrease of the shear modulus with
temperature  observed in experimental  
studies~\cite{fefferman,haziot,haziot1}. As a first step 
we study the effect of the superfluid field on dislocation motion, i.e. 
we investigate the damping of dislocation line motion. To do so, we  calculate 
the mobility of a gliding dislocation line, which in a conventional crystal 
corresponds to 
an inverse viscosity,  in a
quantum crystal. Our calculation of the dislocation motion mobility
is performed  by extending a well known procedure developed in earlier work 
on quasicrystals~\cite{Luben} to quantum 
crystals. We use a %sull4 use
straightforward hydrodynamic approach: %sull3 broke sentence, typo
the hydrodynamic equations 
for $^4$He crystals as 
developed in Refs.~\cite{Andreev,Saslow}. 
Taking into
account  earlier studies~\cite{suzuki,suzuki1} on the role that dislocation lines 
play in 
determining the elastic properties of solid $^4$He crystal, 
we  relate the contribution to the 
dislocation mobility produced by 
the assumed field, to the shear modulus of the crystal.

Obviously, this would be a purely abstract exercise
unless we made contact with experiment: we can claim
validity for 
our ideas only after showing that
they provide at least an alternative explanation for
known experimental facts. Therefore
we  then model existing shear modulus experimental 
data,~\cite{fefferman,haziot,haziot1} 
and show that the drop in shear modulus
with increasing temperature is perfectly consistent with
the existence of an underlying rapid increase of the superfluid 
order parameter as the temperature  decreases. 
Our assumption of a superfluid field associated,
not with bulk superfluidity but
rather  with the %cdsull moved reworded
dislocation line cores, does not
violate, as we shall see below, the present experimental upper limit on the
possible superfluid fraction.
We conclude that the damping of dislocation motion due to 
superfluid field is 
important, in the low temperature limit,  
to be considered with other sources of damping such as $^3$He %sull3
impurities
and thermal phonon scattering. %\hl{It should be noted that we do not  %cdsull on experimental upper limit 
%assume the existence of bulk superfluidity and as we will see below 
%our assumption of superfluidity near the cores of dislocation lines
%does not violate the present experimental upper limit on the
%superfluid fraction. }
 
The experiments~\cite{fefferman,haziot,haziot1} on the shear modulus 
behavior of solid $^4$He
use a procedure where a shear strain 
is applied on the crystal at a set frequency, $\omega$. 
At the  lower frequencies %otvrev
utilized in experimental studies (as low as 2 Hz for the applied strain), 
the experiments are best analyzed, in agreement with
the arguments presented in Ref.~\cite{gorman}, in the 
limit, where the inertial mass of the dislocation 
line can be ignored and one can,
moreover, focus on the quasi-static limit~\cite{jday} for the 
strain due to dislocation motion.  
At higher frequencies the possibility 
arises of a phase shift between the 
applied strain and the resultant displacement of the dislocation line. 
Although  
for the purposes of our study, we focus mainly 
on understanding the mechanism behind the sharp drop in shear modulus 
as a function of temperature for a gliding dislocation line, 
our model is amenable also to a higher frequency scenario 
where the dislocation executes some combination of
gliding and string-like 
vibration, out of phase with the applied strain. 
Our procedure enables us, by making
use of  our results on dislocation pinning due to the superfluid field 
present within crystal $^4$He, to model the low frequency shear modulus
temperature behavior seen~\cite{day,day1,day2,fefferman} in  strain experiments
as well as that~\cite{fefferman,day2} of the $Q$ factor at higher
frequencies. 
 
The organization of the rest  of this paper is as follows. 
In section \ref{methods} we 
present the hydrodynamic method used to calculate the mobility 
of a dislocation in a quantum solid,
 and show how the mobility
can be related to the shear modulus. 
In section \ref{experiment} we illustrate how our results can be used to 
model the experimental data on shear modulus softening in response to
increasing temperature. %otvrev
Two figures illustrate the comparison between theory and experiment: very
satisfactory agreement is found.
Finally, section \ref{summ} contains a summary of the main results and
concluding remarks.

\section{Methods}
\label{methods}
\subsection{Dislocation Mobility}
\label{dmcalculation}

We begin here by describing
the method we use to calculate the dislocation mobility 
in a quantum crystal. This is based on an extension
of the procedure~\cite{Luben}
previously employed to compute the dislocation mobility in
quasicrystals, combined with the 
usual hydrodynamic equations~\cite{Andreev,Saslow} for quantum
crystals. 

%otvam inconsistencies in arrow V_D notation removed in many places below
%otvch changes edited
The considerations below are valid for both edge and screw dislocations. 
For illustrative  purposes only, it is 
helpful to focus on edge dislocations. %abdch 
Consider such a dislocation %otvch
with Burgers vector of length $b$ in a crystal 
subjected to a  
shear stress $\sigma$. 
The shear stress will result in a force per unit length 
on the dislocation line, $\vec{F}_{D}$, which will
cause the dislocation line to move. 
The velocity($\vec{V}_{D}$) of the dislocation line~\cite{Luben} is then proportional 
to $\vec{F}_{D}$: 
%and is related by
\begin{equation}
\label{vd}
\vec{V}_{D} = M\vec{F}_{D}
\end{equation}
where $M$ is, by definition, the mobility coefficient
and, for simplicity, we  
consider the vectors $\vec{F}_{D}$ and $\vec{V}_{D}$ to be parallel
to each other. %otvam moved %%We will later %otvr
%call the direction of these vectors 
%the $x$ direction. 
The general approach seen in~\cite{Luben} and~\cite{Bakai}  to 
the calculation of  the mobility
involves equating the rate of work done by a force applied on the dislocation 
line, which causes the line to move
with a constant speed $V_D$, to the 
energy dissipation rate due to the fields associated with the dislocation 
line motion.
Thus, one has, 
\begin{equation}
\label{fv}
{F}_{D}{V}_{D} = -\frac{d}{dt} \int d^{2}r E_{el}, %otvam arros removed
\end{equation}
where the right hand side is %otvam , as stated above,
the rate at which energy is dissipated in the 
elastic fields, as mentioned above, and the integral
is over a two dimensional plane orthogonal to the 
%straight edge dislocation.
dislocation line. %otvch 
The 
left hand side
is the rate at which energy is transferred onto the dislocation line due to the 
applied force $F_{D}$. 

To calculate the mobility, one isolates the terms in the right
side of Eq.~(\ref{fv})
that are proportional to 
$V_{D}^2$. Then one extracts $M$ from the relation:
\begin{equation}
\label{inv}
F_{D}V_{D} =M^{-1}V_D^2, %abd deleted V^2 and also moved the statement regarding direction of vectors
\end{equation}
where we have made use of Eq.~(\ref{vd}). %otvam and, %otvr 
 %otvr
The left side
of Eq.~(\ref{inv}) is evaluated from the right side of Eq.~(\ref{fv})
using hydrodynamic methods described below. 
Then, the constant of proportionality 
between the dissipation and the square
of $V_D$ is the inverse mobility of the 
dislocation line. 
 
The contribution to the mobility of the dislocation 
line that we wish to calculate arises from physical
phenomena   at
length scales much larger than the dislocation core size. 
%abdch %otvch
A hydrodynamic description of superfluidity  confined to narrow channels
is valid whenever~\cite{cdotv} the width of such channels exceeds
the coherence length. Hence, it is valid in our case 
%For narrow superfluid channels associated with dislocation lines the 
%validity of the hydrodynamic method used here maybe questioned. However, 
%A hydrodynamic 
%description is valid~\cite{cdotv} 
because  the coherence length is of the order of interatomic
distance, and the size  
of  the typical superfluid region near the core of a dislocation 
line is~\cite{troyer,troyer1} 
of order $\sim$10 times the interparticle spacing or more. %otvch til here
%abdch till here
Hence,  we can use the hydrodynamic methods of Ref.~\cite{Luben}.
%otvch eliminate repetion? which are valid at such length scales.  %abdch1 agree
The proper hydrodynamic  approach in our case 
is as developed in Refs.~\cite{Andreev,Saslow}.  
We particularly follow the notation of the latter.

We have for elastic energy density $E_{el}$ and its
differential, $dE_{el}$, the expressions:
%and the energy density, and the associated Gibbs-Duhem relation
%\begin{eqnarray}
\begin{eqnarray}
\label{ener}
dE_{el} &= T ds + \lambda_{ik} dw_{ik} + \phi d\rho + \vec{v}_{n}.d\vec{g} + 
\vec{j}_{s}.d\vec{v}_{s}, \nonumber \\
E_{el} &= -P + Ts + \lambda_{ik} w_{ik} + \phi \rho + \vec{v}_{n}.\vec{g} + \vec{j}_{s}.\vec{v}_{s},
%\end{eqnarray}
\end{eqnarray}
with the associated Gibbs-Duhem equation:
\begin{equation}
\label{gd}
0 = - dP + s dT + w_{ik} d\lambda_{ik} + \rho d\phi + \vec{g}.d\vec{v}_{n} + 
\vec{v}_{s}.d\vec{j}_{s}.
\end{equation}
In these expressions, $T$ is the temperature, $s$ is the entropy density, 
$P$ the pressure, $\lambda_{ik}$  the elastic tensor
density, $\phi$  the chemical potential, $\rho$ the mass density, 
$\vec{v}_{n}$ the normal fluid velocity,
$\vec{g}$  the momentum density, $\vec{j}_{s}$ is the superfluid momentum 
density, and $w_{ik}$ is the 
%non symmetric %abd3 calling it the strain tensor
strain tensor defined as, 
\begin{equation}
\label{wik}
w_{ik} = \partial_{i}u_{k},
\end{equation}
associated with the lattice displacement $\vec{u}$. %otvam 
The subscripts $n, s$ denote normal and superfluid components respectively
while $i,j,k$ are coordinate indices. Here and in the rest of the paper
summation over repeated coordinate indices is implied. %otvam 

The linearized hydrodynamic equations of motion~\cite{Saslow} are: 
%\begin{subequations}
\begin{eqnarray}
%\label{hydro}
%\begin{eqnarray}
\label{hydro}
\partial_{t} \rho + \partial_{i}g_{i} & = & 0, \nonumber \\
\partial_{t} g_{i} + \partial_{k}\Pi_{ik} & = & 0, \nonumber \\
\partial_{t} s + \partial_{i} f_{i} & = & -\frac{q_{i}}{T^2} \partial_{i}T ,
\nonumber \\
\partial_{t} v_{si} +  \partial_{i} \phi & = & 0, \nonumber \\
\partial_{t} u_{i} & = & v_{ni}.
%\end{eqnarray}
\end{eqnarray}
where 
%\begin{equation}
\begin{eqnarray}
\label{def}
g_{i} & = & \rho_{sik} (v_{sk} - v_{nk}) + \rho v_{ni}, \nonumber \\
\Pi_{ik} & = & -\eta_{iklm}\partial_{m}v_{nl} -\zeta_{ik} \partial_{l}j_{sl} + P\delta_{ik} - \lambda_{ki},
\nonumber  \\
f_{i} & = & sv_{ni} + \frac{q_{i}}{T}, \nonumber \\
\phi & = & -v_{sk}v_{nk} + \zeta_{ik} \partial_{k}v_{ni} + \chi\partial_{k}j_{sk}.
\end{eqnarray}
%\end{equation}
These expressions can be found, with one minor~\cite{wayne} difference
in Ref.~\cite{Saslow}.
%Talk about shear viscosity vs second viscosity here
In the expression for the momentum flux tensor ($\Pi_{ik}$) above, 
the tensor with components $\eta_{iklm} $ is a shear viscosity 
arising from the normal component. On the other
hand, the tensor $\zeta_{ik}$ and 
the scalar $\chi$ are  ``second viscosity'' %otvam 
coefficients arising from coupling between normal and superfluid
components. Also, $f_{i}$ is the entropy flux and $q_i$ is the thermal current.
 
Having established a framework for calculating the energy  %otvam
dissipation associated with 
dislocation motion, we now turn to the evaluation of 
the strain term contribution to the %otvam
 mobility. 

\subsection{Contribution of the strain term to the mobility}

We now explain the procedure used to calculate the
dislocation mobility. 
We illustrate the details by considering first the 
contribution to $M$ arising from the 
energy dissipation $E_{strain}$
associated with the strain field ($w_{ik}$) of 
the dislocation line. Other contributions will be discussed later.
The expression for this source of energy dissipation  is obtained by making use of
Eqns.~(\ref{ener}), (\ref{wik}) and (\ref{def}):
\begin{eqnarray}
\label{edot}
%\begin{eqnarray}
\dot{E}_{strain} & = & \lambda_{ik} \dot{w}_{ik}\nonumber \\
& = & (-\eta_{kilm}\partial_{m}v_{nl} -\zeta_{ki} \partial_{l}j_{sl} + P\delta_{ki} -\Pi_{ki})
\partial_{i} \dot{u}_{k},
\end{eqnarray}
%\end{equation}
where the overdot denotes the time derivative. %otvam
%based on the definition of $\lambda_{ik}$ in Eq.(\ref{def}) and Eq.(\ref{wik}). 
Following Ref.~\cite{Luben}, $\vec{u}$ is the displacement field of 
the crystal lattice sites from their
equilibrium positions due to a dislocation line moving with
constant velocity $\vec{V}_D$. Thus, we assume the space and time
dependence of $\vec{u}$ to be of the form
\begin{equation}
\label{ud}
\vec{u}(\vec{r},t) = \vec{u}(\vec{r} - \vec{V}_{D}t),
\end{equation}
which implies 
\begin{equation}
\partial_{t} \vec{u} = -(\vec{V}_{D}\cdot\vec{\nabla})\vec{u}
\end{equation}
corresponding to the velocity of atoms moving with the dislocation line. %abd added line
Also, %abd3 grammar
$\vec{v}_{s} = \partial_t\vec{u}_{s}(\vec{r} - \vec{V}_{D}t)$, is the velocity of   
superfluid atoms due to the moving dislocation line. 
Keeping the relevant dissipative terms in 
$\dot{E}_{strain}$, which lead to a $V_{D}^2$ dependence 
(for example in the term $\eta_{kilm}\partial_{m}v_{nl} \partial_{i} 
\dot{u}_{k}$ both $v_{nl}$ and $\dot{u}_{k}$ depend on $V_{D}$), we have:
\begin{equation}
\lambda_{ik} \dot{w}_{ik} = (-\eta_{kilm}\partial_{m}v_{nl} -\zeta_{ki} \partial_{l}j_{sl})\partial_{i} 
\dot{u}_{k}
\end{equation}
where,  $j_{sl} = \rho_{slk} v_{sk}$ and $v_{n}$ are the supercurrent
and the normal fluid velocity.  
Making then use of Eq.~(\ref{fv}), we
have for the energy dissipated in the strain field of the dislocation 
line:
%\begin{equation}
\begin{eqnarray}
\label{estrain}
F_{D}V_{D}\vert_{\eta,\zeta} & = & -\int \dot{E}_{strain} d^{2}r, \nonumber \\
& = & \int (\eta_{kilm}\partial_{m}v_{nl}\partial_{i}\dot{u}_{k} + \zeta_{ki} \partial_{l}j_{sl}
\partial_{i}\dot{u}_{k})d^{2}r,
\end{eqnarray}
%\end{equation}
where the notation in the left side denotes the contribution
from the $\eta$ and $\zeta$ tensors, under examination.
We now establish a coordinate system with the $z$ axis directed along the 
dislocation line and %otvam
the $x$ axis along the 
velocity. We then perform
a two dimensional Fourier transform for the displacement field in the transverse
directions: 
\begin{equation}
\label{fourier}
\vec{u}(\vec{r}) = \int \vec{u}(\vec{q}) e^{i\vec{q}.\vec{r}} d^{2}q.
\end{equation}
%abd3 editing below
We illustrate below the steps involved in
evaluating  the inverse dislocation mobility 
contribution due to the shear viscosity term $\eta_{kilm}$
in Eq.~(\ref{estrain}) above. The inverse mobility contribution
 from the term $\zeta$
is evaluated in a very similar way. %abd moved this 
The next steps, then, involve collecting the contributions
from different dissipative coefficients in Eqs.~(\ref{ener}) and (\ref{def}). 
%abd3 till here %otvam no new para
Inserting the Fourier transform of the displacement field as in 
Eq.(\ref{fourier}) into 
Eq.(\ref{estrain}) (also using the last of Eq.(\ref{hydro})) we obtain
\begin{eqnarray} %abdiop left align
%abdjltp deleted \fl used for iop does not work
%abdjltp adjusted to fit the equation within the boundary
\label{calc}
&&F_{D}V_{D}\vert_\eta =   \int(\eta_{kilm}\partial_{m}v_{nl}\partial_{i}\dot{u}_{k})d^{2}r \nonumber  \\
 &=&\int [\eta_{kilm}\partial_{m}(\vec{V}_{D}\cdot\vec{\nabla})  
\int u_{l}(\vec{q}_1)e^{i\vec{q_1}\cdot\vec{r}}d^{2}q_1 \partial_{i}(\vec{V}_{D}\cdot\vec{\nabla}) 
 \int u_{k}(\vec{q}_2)e^{i\vec{q_2}\cdot\vec{r}}d^{2}q_2] d^{2}r \nonumber \\ %abd3 adding bracket for clarity
&=& \eta_{kilm} \int \int iq_{1m}(i\vec{V}_{D}\cdot \vec{q}_1)u_{l}(\vec{q}_1)e^{i\vec{q_1}\cdot
\vec{r}}d^{2}q_1\times \nonumber \\
&& \int iq_{2i}(i\vec{V}_{D}\cdot\vec{q}_2) u_{k}(\vec{q}_2)e^{i\vec{q_2}\cdot\vec{r}}d^{2}q_2 d^{2}r. 
\end{eqnarray}
Noting that the integral over the 2D plane leads to a two-dimensional
$\delta(\vec{q_1}+\vec{q_2})$ and using this delta function to 
%\begin{equation}
%\int e^{i(\vec{q_1}+\vec{q_2}).\vec{r}} d^{2}r = \delta(\vec{q_1}+\vec{q_2}),
%\end{equation}
integrate over $\vec{q_2}$ %using the delta function above 
one then obtains
\begin{equation}
F_{D}V_{D}\vert_\eta = \eta_{kilm} \int q_{1m} q_{1i}u_{l}(\vec{q}_1)u_{k}(-\vec{q}_1)(V_{D}^2 q_{1x}^2) d^{2}q_{1}.
\end{equation}

To estimate this contribution to the mobility it is
sufficient to consider  the diagonal component of the viscosity,  
%viscosity tensor 
%reduces to
the ordinary %otvr
 $\eta \equiv 
\eta_{iiii}$. Then, the rate of energy loss in the strain field of a 
dislocation line simplifies to: 
\begin{equation}
\label{fdeta}
F_{D}V_{D}\vert_\eta = \eta \int q_{1i} q_{1i}u_{i}(\vec{q}_1)u_{i}(-\vec{q}_1)(V_{D}^2 q_{1x}^2) d^{2}q_1.
\end{equation}
By simplifying the equation above and finding the inverse mobility of the dislocation line 
as explained after Eq.(\ref{inv}) we obtain for this contribution to $M^{-1}$: %abd3 minor typo
\begin{equation}
\label{mobeq}
M^{-1}\vert_\eta = \eta \int_{q_{min}}^{q_{max}} q_{1i}^2 |u_{i}(\vec{q}_1)|^2 q_{1x}^2 d^{2}q_1 %abd3 adding index i
\end{equation}
where $q_{min} = 1/L$, $L$ being a cutoff of order
of the size of the crystal, or the distance between dislocations,
and $q_{max} = 1/b$   
with $b$ (the magnitude of Burgers vector) approximately comparable to the interatomic spacing.
To obtain $\vec{u}(\vec{q})$, (the %otvam
Fourier transform of the elastic displacement field)
we note that the gradient of the elastic displacement field is 
roughly  constant in magnitude over a 
circular path of radius $r$ 
centered on the dislocation line~\cite{Luben} i.e.
\begin{equation}
(\nabla u) r \approx b.
\end{equation}
Fourier transforming the elastic displacement field then leads to:
\begin{equation}
\label{uq}
u(\vec{q}) \approx \frac{b}{q^2}.
\end{equation} 
Substituting Eq.~(\ref{uq}) into Eq.~(\ref{mobeq}) we finally have:
%\begin{equation}
\begin{equation}
M^{-1}\vert_\eta = \frac{\eta}{2} (1-\frac{b^2}{L^2})
           \sim \frac{\eta}{2}.
\end{equation}
%\end{equation}
since obviously $b \ll L$.

As mentioned above, in
addition to this term arising from $\eta$, which we have
discussed in detail, and the corresponding contribution
from  the
$\zeta$ tensor, written down in Eq.~(\ref{estrain}) and computed in the same way %abd3 changed eq reference from ref{edot}
as the $\eta$ term,
there are several additional
contributions to the dissipation
%In taking into account the various contributions to energy dissipation
all arising from terms in Eq.~(\ref{ener}). We 
can ignore the contribution from the $T\dot{S}$ 
because we 
are interested in the limit where $T \rightarrow 0$. Also, since we are 
interested in the limit 
where $V_{D}$ is small compared to 
the speed of sound (in solid $^4$He) the 
inertial contribution
to energy dissipation, $\vec{v}_{n}\cdot\dot{\vec{g}}$, can be 
neglected~\cite{Luben}. There are additional  
%Using a similar procedure as noted above, other 
contributions to the mobility of the dislocation line  arising
from the $\phi \dot{\rho} + \vec{j}_{s}\cdot\dot{\vec{v}}_{s}$ terms
in Eq.~(\ref{ener}). These involve again the tensor $\zeta$, this time
via the last of Eqs.~(\ref{def}) and also, via the same equation, the
scalar $\chi$. These contributions 
can be calculated following similar procedures to those discussed
above and there is no need to repeat the details. 
Finally, putting together  
all of these contributions, the total inverse 
mobility 
of the dislocation line in a quantum crystal can be written as
\begin{equation}
\label{mobility}
M^{-1} \approx \frac{\eta}{2} + \frac{\rho_s}{\rho}(\zeta \rho + \chi \rho^2).
%otvam
\end{equation}
This is our basic result for the mobility.

\subsection{Relation between shear modulus  and  mobility} %otv2 edited
\label{relation}
We will now 
proceed further to
relate the observed shear modulus of a 
quantum crystal to the mobility of a dislocation line. This will enable us to
discuss the experimental results of Ref.~\cite{fefferman}. %abdrev1 typo 
%abd3 notation change from small d to D for dislocation 

When a stress $\sigma$ is applied to a crystal, the dislocation line 
%abd1 adding content
feels a force per unit length, $F_D = b\sigma$.
As the dislocation line glides in response to the applied force,
the displacement of the dislocation line results of course in a strain 
$\epsilon_{D}$ in addition to the strain $\epsilon_{el}$. Here the elastic 
strain $\epsilon_{el}$ is 
the response of the crystal in the absence of dislocation lines. 
Therefore, the effective shear modulus $\mu$ i.e. 
the ratio of applied stress to total strain is given by~\cite{fefferman}
 $\mu = \sigma/ (\epsilon_{D}+\epsilon_{el})$. This 
can also be written as
\begin{equation}
\mu = \frac{\mu_{el}}{1+\frac{\epsilon_{D}}{\epsilon_{el}}}, 
\end{equation}
where $\mu_{el} = \sigma/\epsilon_{el}$ is the elastic shear modulus. 
The strain due to the motion of the dislocation line is known to 
be~\cite{orowan,taylor}
\begin{equation}
\label{strain}
\epsilon_D = \rho_D b x_a, %abdrev replacing bar x by x %otvrev x_a
\end{equation}
where $\rho_D$ is the dislocation number density and 
$x_a$ is the average displacement 
of the dislocation line through the crystal. 
In order to determine the dislocation displacement, 
we look to the applicable equation of motion~\cite{granato} 
of the dislocation line
\begin{equation}
\label{displ}
M^{-1}\dot{x} = b \sigma. 
\end{equation}
where $x$ is the dislocation displacement as a function of time. %otvrev
%abdrev further edits to \tau vs \omega below %otvrev moved to later
%and position between the pinning points of the dislocation line 
%and $M^{-1}$ is the inverse mobility. 
We can then relate the average  displacement  of the dislocation 
to its velocity %abdrev1 amplitude to avg
and to
mobility via $x_a = V_D \tau$  %otvrev
(where $\tau$ is the 
characteristic time scale associated with 
the movement of the dislocation line) 
and  Eq.~(\ref{vd})
to obtain $\epsilon_D=F_D\rho_DbM \tau$, which agrees
with Eq.~(\ref{displ}) and $F_D=b\sigma$. 
For the purposes of %otvamf
estimating the the average displacement  $x_a$, we will, working %otvrev edits
as in Ref.~\cite{jday} in the quasistatic limit, replace $\tau$ by the 
inverse 
of the slowest range of frequencies for the
strain $\epsilon$ applied to the crystal 
in the experimental~\cite{fefferman}
situation.  

Putting together these 
considerations, the shear modulus of the crystal as a function of the 
mobility of the dislocation line is found to be
\begin{equation}
\label{muresult}
\mu = \frac{\mu_{el}}{1+\frac{F_{D} \rho_{D} b M \tau}{\epsilon_{el}}} . 
\end{equation}
%where we have used the definition Eq.(\ref{vd}). 
We will %otvam
now be able to use  this result,
combined with that for the mobility in the previous subsection
(Eq.~(\ref{mobility})) to discuss the behavior of the elastic coefficient
in solid Helium. 

The expressions derived above %abdrev quasistatic
for the amplitude of displacement and $\epsilon_D$ are valid 
under the standard~\cite{gorman,jday} %otvrev %abdrev1 adding one more quasistatic ref. 
quasistatic limit assumptions.  
%{\it in phase} with the applied strain. %otvamf
This  is applicable in the  experimental
situation since 
the dislocation acceleration time is small
compared to the time over which strain is applied 
to the crystal~\cite{note}. However, %otvamf reworded
%abd2 AC limit
we can also consider the   higher frequency 
limit (the ``ac'' limit) 
when the dislocation line 
executes elastic string motion out of phase with the applied strain. 
%(which we will refer to as the ac limit) applicable at sufficiently high frequencies. 
%In this limit, we model dislocation lines 
%as elastic strings that bow as a result of stress. 
Considering then, 
$x(t) = x_0 e^{-i\omega t}$ and $\sigma(t) = \sigma_0 e^{-i\omega t}$ 
%otvrev and taking into account the  phase difference of $\pi/2$ between %otvamf
%otvrev stress and strain in this limit, %otvamf
one obtains 
from Eq.(\ref{displ}):
\begin{equation}
x_0 = \frac{i b \sigma_0}{M^{-1} \omega}.
\end{equation}
Following then the same steps as in the previous case  
we can easily find $\mu(\omega)$ in this ``ac'' limit. %abdrev1 omega
The quantity of interest here is the $Q$ factor 
$\left|Q^{-1}\right| = | Im[\mu]/Re[\mu]|$.
Defining 
$\epsilon_R(\omega) \equiv \frac{\rho_D b F_D M * (1/\omega)}{\epsilon_{el}}$ as 
a ratio involving strains,
one finds   that $Im[\mu]=-\epsilon_R Re[\mu]$,
leading to
the  expression for the $Q$ factor,
 $\left|Q^{-1}\right| = | Im[\mu]/Re[\mu]  | = \epsilon_R(\omega)$. 

\section{Results - Modeling of Experimental Shear Modulus Data} %otv2 edited
\label{experiment}
%-	Describe how the superfluid contribution to mobility is much larger than the other contributions in the T -> 0 limit. 
%-	Explain how this new mobility effect can model the anomalous shear modulus behavior of solid 4He. 
Having derived the mobility of the dislocation line and its relation 
to the shear modulus of a quantum crystal, 
we now discuss how to connect our theory to the experimentally observed
large and sudden softening of the shear modulus seen~\cite{fefferman} in 
solid $^4$He as the temperature is increased. 
%the connection between dislocation mobility and shear modulus behavior in solid 
%$^4$He crystals. 
%Using the mobility results obtained above, 
Thus, we will seek to model the 
temperature dependence of the shear modulus data.
%in the experiment mentioned above.
%\cite{fefferman} on  the shear modulus of solid $^4$He crystals. 
Results in Ref.~\cite{fefferman} show that at the higher
temperatures studied, (up to 1K), the crystal 
is softer, with $\mu$ being independent of temperature. As the temperature 
is lowered, it is seen that
between $T=$ 0.05K and 0.1K 
%depending on the frequency at which shear stress is applied,
$\mu$ rises sharply, and then, at
lower temperatures, it 
saturates to a value identified
with the  intrinsic value $\mu_{el}$. 
It is evident from earlier studies~\cite{suzuki,suzuki1} that dislocation lines play an important role
in determining the elastic properties such as shear modulus of a crystal.  
 Pinning due to 
impurity atoms and collision with thermal phonons have
been  considered to be the dominant source for
damping of dislocation line motion in recent experiments~\cite{haziot,haziot1}: %CD shifted citation
%we drop the contribution from $\eta$ in favor of the 
hence we have to compare these sources of damping with the 
superfluid contribution
in order to ascertain their relative importance. 
In $^4$He crystals, dislocation lines can glide almost freely along 
the basal planes of the hexagonal crystal structure~\cite{haziot}.  
As a consequence, the effect of the shear viscosity (i.e. 
the dissipation due to $\eta$, arising from 
interactions between the dislocation line and 
the surrounding atoms) can be neglected.
It follows from the above argument that the 
expression for the dislocation mobility, Eq.~(\ref{mobility}) then simplifies
to: 
\begin{equation}
\label{simpleM}
M^{-1} \approx \frac{\rho_s}{\rho}(\zeta \rho + \chi \rho^2).
\end{equation}
An important consequence of this result is that a
superfluid field makes it harder for the dislocation line to move. 
Since the inverse mobility is 
directly proportional to the superfluid fraction, at lower temperatures (as 
$T \rightarrow 0$) when we expect $\rho_s/\rho$ to be larger, it is 
harder for the dislocation line to move. 
Even though this appears to be counterintuitive, %otvr better explanation needed
it can be understood by recalling that %otvr vibrating wire
%otvr experiments do find that superfluid $^4$He is characterized by a shear viscosity even though it is very small. 
%abd2 explaining better.
Couette viscometer~\cite{woods} and vibrating wire~\cite{tough} experiments show that liquid $^4$He
is characterized by a very small viscosity of order $10^{-5}$ kg/ms below 1K. 
This is relevant because the small viscosity contribution from the superfluid field 
could be the dominant source of dissipation for a dislocation line in a quantum solid. 
%otvr abdul why is this relevant? %abd2 clarifying

We can now numerically estimate the inverse mobility of the dislocation line. 
To do so, we will use
the values of the second viscosity coefficients 
$\rho \zeta \sim 10^{-5}$ kg/ms and $\rho^2 \chi \sim 7*10^{-5}$ kg/ms
for liquid $^4$He~\cite{putt,um}. 
These coefficients are not known  for solid $^4$He. 
We will then obtain  corresponding
estimates for the dislocation mobility
both above and below the assumed superfluid transition temperature 
(or crossover) in solid 
$^4$He crystals. %abdrev editing transition temp discussion below
For the purposes of estimating the %otvrev edited superfluid transition 
temperature  at which a superfluid field associated
with dislocations may arise 
in solid $^4$He crystals,
we consider a scenario where a loosely intersecting 
grid of dislocations forms. 
According to Refs.~\cite{shev,aleinikava} dislocation network superfluidity
is characterized by two temperature scales - $T_0 \sim$ 1K 
(comparable to the bulk $\lambda$ temperature for liquid $^4$He) and 
$T_c \sim T_0 a/L_f$, where $a$ is the interatomic distance along 
a dislocation core 
and $L_f$ is the free segment length of dislocation line.
Within such a model for superfluidity associated 
with dislocation lines, 
the onset temperatures are roughly
consistent with  %abdrev1 would to could and or smaller wording below
the experimentally~\cite{fefferman,chan,chan1} found range 0.1K to 0.075K  where 
the onset of anomalous %abdrev2 changed 0.25K to 0.1
behavior is observed. 
As to the numerical value of  the putative superfluid fraction 
$\rho_s/\rho$ we look to quantum Monte Carlo calculations involving 
solid $^4$He. %abdiop deleted sentence
%For solid $^4$He in the hcp lattice structure,  %otvrev
%the superfluid fraction ($\rho_s/\rho$)
%was calculated to have an upper limit in the range of 0.3 to 0.9.~\cite{Saslow}
Monte Carlo simulations with dislocations characterized by core
superfluidity find that nearly all atoms in the core of a dislocation line
are in the superfluid state~\cite{troyer}. We will conservatively
take the value $\rho_s/\rho \sim 0.1$ 
in the $T \rightarrow 0$ limit 
%abdiop adding sentences below
for the superfluid fraction near the dislocation core.
%otviop 
%and not averaged over the whole sample. 
It is very easy to see that for
any reasonable values of the dislocation density
in solid $^4$He crystals~\cite{rojas}, 
and 
of the radius from the dislocation core in which superfluidity 
might be present, this estimate of the superfluid density near the core
leads to a superfluid 
fraction averaged over the whole sample well below the current
experimental upper limit for $\rho_s/\rho \sim 10^{-6}$~\cite{kim2}. 
%otviop reference order fixed
%abdch 
%otvch some edit
Monte Carlo studies of superfluidity associated with dislocation lines 
focus on whether a fully connected three-dimensional network of 
superfluid flow is established as a possible explanation to the purported
observation of period drop in torsion oscillator experiments. %abdch1 
In this context, it is found~\cite{proko}  
that edge dislocations can display finite  superfluid response.
Other simulations~\cite{soyler} 
consider 
a temperature range (0.5 K) much higher
than what is relevant here.   
%much higher than the temperature range of interest in the current paper.
% otvch end

%abdch till here

Using these order of magnitude values we can obtain,
via Eq.~(\ref{simpleM}), an approximate
value for the inverse mobility ($M^{-1}$)
at low $T$  namely
$\sim 10^{-5}$ kg/ms. 
In some of the previous work,~\cite{haziot2}
damping of 
dislocation motion was thought to be mainly due to phonon collisions and 
pinning effect due to the presence of $^3$He impurities,
%The inverse mobility of the dislocation line due to phonon scattering 
%and pinning due to impurities are 
resulting in inverse mobilities of order 
$10^{-9}$kg/ms and $10^{-8}$kg/ms, respectively, at 1K.
Hence, our results 
indicate that quenching of dislocation motion due to superfluid field may be 
the dominant source of damping for dislocation motion in the low temperature limit.
In this limit, the superfluid contribution to dislocation damping is 
larger by at two  or three orders of magnitude compared to the other sources of damping.
%considered in recent experiments. ~\cite{haziot2}

%otvrev %abdrev below is new
In order to estimate the magnitude of  the 
shear modulus using Eq.~(\ref{muresult}), (with Eq.~(\ref{simpleM})),
we need also the relevant 
values of the other parameters entering 
that equation.  From experimental 
results~\cite{haziot,fefferman,rojas}, we have %CD shifted citation
$F_{D} \sim 10^{-11} N/m$,  %abdrev added reference for \rho_D
$\rho_{D} \sim 10^{6} m^{-2}$, $b \sim 10^{-10} m$ and 
$\epsilon_{el} \sim 10^{-8}$. As mentioned above, we use 
the inverse of the applied frequency, in the low range,
$\omega \sim$ 1 Hz - 50 Hz, to estimate 
$\tau$. This should be better at lower
frequencies. Inserting the values of the various parameters 
into Eq.~(\ref{muresult}) and taking $M^{-1} \sim 10^{-5}$ kg/ms
in the low $T$ limit as mentioned
above, we obtain the ratio
in the denominator of that equation to be 
\begin{equation}
\label{mulowt}
\frac{F_{D} \rho_{D} b M \tau}{ \epsilon_{el}} \sim 10^{-3} - 10^{-5}.
\end{equation}
%in the low temperature limit. 
It can then be easily seen from Eq.~(\ref{muresult}) that 
$\mu \approx \mu_{el}$ in the low $T$ regime. 
%These  were obtained 
%abdrev including new figure and including description
\begin{figure}
\begin{turn}{-90}
\includegraphics[width=0.65\textwidth] {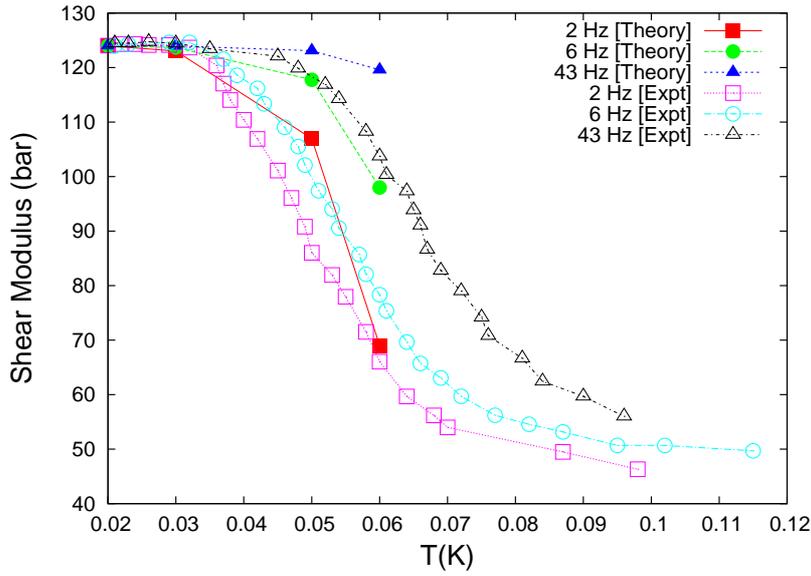} %oa
\end{turn}
\caption{(Color online) Experimental %otvrev
results from Ref.~\cite{fefferman}
for the shear modulus vs temperature at different 
frequencies, $\omega$, of applied strain
(see legend) are 
compared with theoretical results. See text for discussion.}
\label{mob}
\end{figure}
Thus, at low temperatures (below $T \sim$ 0.03K), we find that the shear 
modulus becomes independent of $T$ 
at $\mu \approx \mu_{el}$, a result that should not depend on frequency
at low $\omega$. 
In comparing the experimental results for the shear modulus with 
our model (see Fig.\ref{mob}), we note that the low temperature behavior of 
this shear modulus is well accounted for in our theory. 

Examining now the behavior of the dislocation mobility near and above the 
temperature where the
assumed superfluid behavior disappears,
we note that the contribution to the inverse mobility due 
to the superfluid field approaches zero
as $\rho_s/\rho \rightarrow 0$. Then, other damping effects such as 
pinning due to $^3$He impurities and phonon 
scattering should %otvrev %abdrev1 adding 3He and changing order
become more important and eventually
dominant. The experimental
range of temperatures where the sharp change in shear
modulus is found in solid $^4$He (as mentioned above), 
%the temperature at which the
%drop in shear modulus is experimentally observed 
is accounted for in our theory. 
In order to estimate the  shear modulus in the  
higher temperature 
range (T $\geq$ 0.03K) we first note~\cite{fefferman,haziot} %otvrev reference m 
that $M^{-1} \sim 10^{-8}$ kg/ms at $T=0.06K$.
This implies that the ratio in Eq.~(\ref{mulowt}) is of order unity and
$\mu \approx 0.5 \mu_{el}$. At the intermediate value $T=0.05 K$ we
interpolate $M^{-1} \sim 5*10^{-7}$ kg/ms. At lower $T$ the precise value
of $M^{-1}$ becomes irrelevant. The results thus obtained are displayed
in Fig.~\ref{mob}. 
%
%we approximate %abd3 minor correction %abdrev 
%$M^{-1} \sim 10^{-7}$ kg/ms at 0.03K, 
 %at 0.05K and . 
Other parameter values used there are, except of course for
$\rho_s$, the same as in the low temperature range. 
As shown there, the 
drop in shear modulus is modeled quite well by Eq.~(\ref{muresult}) %abdrev1 typo 
and the above considerations, %otvrev 
specially at the lowest frequencies. In the higher frequency 
range considered the quasistatic
approximations might start to break down.
We also note that at a given temperature, as the 
frequency $\omega$ is lowered 
the value of the shear modulus decreases in agreement with
experimental data. 

\begin{figure}
\begin{turn}{-90}
\includegraphics[width=0.65\textwidth] {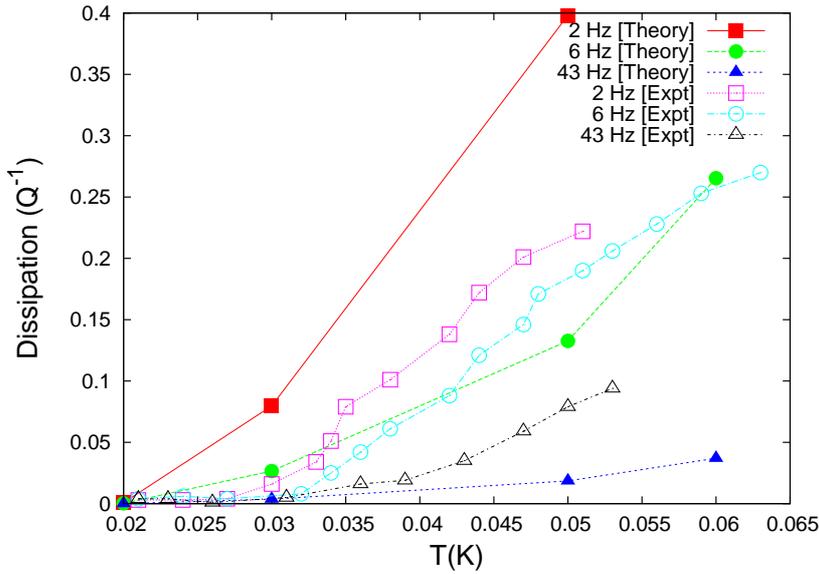} %oa
\end{turn} %otvrev
\caption{(Color online) Dissipation ($Q^{-1}$) vs temperature at different 
frequencies, $\omega$, of applied strain
(legend). Experimental data  from Ref.~\cite{fefferman}
are compared with theory, as discussed in the
text.}
\label{diss}
\end{figure}

%abdrev edit till here regarding new figure
%abd2 ac limit discussion %abdrev moving ac discussion here %otvrev edited
Considering now the higher frequency ``ac'' limit,
we focus, as mentioned above, on the $Q$ factor. 
In Fig.~\ref{diss}, we plot the dissipation associated
with dislocation line vibration, $Q^{-1}$ (i.e. 
arising from the phase difference between 
$\sigma$ and $\epsilon$) as obtained from $\left|Q^{-1}\right| =  \epsilon_R$. 
(see discussion at the end of Sec.~\ref{relation}). 
Dissipation is small 
at low temperatures and increases with T. 
The numerical values of the parameters used in calculating
$Q^{-1}$ are the same as for the shear modulus. The experimental
results~\cite{fefferman,haziot1} for $Q^{-1}$ in the low $T$ limit are
again consistent with theory: the disagreement  is now greater at very
low frequencies, as one would expect.
%otvamf 
At temperatures near the superfluid 
onset temperature, we note that experimentally the dissipation 
decreases with larger values of $\omega$.  
 %otvamf  %abdiop adding discussion here %otviop added
This frequency behavior of the dissipation is in
agreement with our results. 

Next, we discuss some of the limitations of our model.  
First, the calculated dislocation
mobility in the hydrodynamic limit provides probably a lower limit 
to this quantity
since we do not take into account the core strain effects of 
the dislocation line.
%taking core effects into account would result in  inverse mobility to be higher.
Also, as noted above, the second
viscosity coefficients used in calculating the mobility are for 
liquid $^4$He rather than for 
solid $^4$He, due to the lack of experimental data in the solid phase. 
Since one might reasonably expect the second viscosity coefficients 
for solid $^4$He to 
be somewhat larger than in the liquid, %abdrev1 edit starting here
it is possible that even for smaller ratios of $\rho_s/\rho$ 
than the value $\rho_s/\rho \sim 0.1$ considered here
the superfluid contribution to the dislocation mobility
could still be significant
compared to the contribution from 
phonon scattering or pinning effects due to impurities. %abdrev1
These reasons, then, might account for the 
remaining differences in %abdrev edit here to account for better agreement. 
the magnitude of %otvrev
the temperature dependence of 
the shear modulus and dissipation as shown in Figs.~\ref{mob} and \ref{diss}. 
%abdiop comment on impurity pinning here %otviop edited
The onset temperature for shear modulus softening was observed to %abdiop1 softening
increase with increasing 
$^3$He impurity concentration~\cite{day}. % with higher impurity 
%concentration leading to higher onset temperature for the drop in shear modulus.  
This behavior can be qualitatively understood from our ideas: 
we have shown that
a model~\cite{malmi} of a dislocation network with quenched disorder
%results of a 
%previous study by the same authors~\cite{malmi} where a 
%quenched dislocation network increases the superfluid
has a higher % superfluid %abdiop1
ordering temperature than in the annealed case. 
Higher impurity concentration would be more effective
in quenching the dislocation network thereby increasing the 
onset temperature of the superfluid field.
%temperature. %otviop moved
More detailed knowledge of the phonon and impurity
effects is thus, we believe,  likely to lead to a more quantitative
agreement in the higher $T$ limit. %otvamf
%otvch below added (OPTIONAL)
In our comparison with experiment, we have implicitly assumed that the
dislocations with superfluid cores couple to the measured shear modulus.
Our mobility estimates can be applied to all dislocations. %abdch1
The arguments given above in the context of estimating the
superfluid fraction and MC results indicate that our assumption
appears justified.
%otvch
%abdch till here

\section{Summary} 
\label{summ}

We have begun, in section \ref{dmcalculation} of this paper, by 
presenting a novel 
calculation of the dislocation mobility in 
solid $^4$He. This calculation is based on well known hydrodynamic
results~\cite{Saslow} and follows a procedure developed in 
Ref.~\cite{Luben}.  
The result is expressed 
in terms of the bulk ``second viscosities''
of the superfluid crystal hydrodynamics, and the value of $\rho_s$,
assumed to be nonzero.
Numerical estimates of the mobility, although rather uncertain,
indicate that a putative superfluid field  associated  with
dislocation lines, can   
play a nontrivial role in dislocation mobility and therefore affect the 
stiffness of the crystal.  
We show that as a  consequence, 
superfluid damping of dislocation motion can model the large 
and sudden increase
in shear modulus observed experimentally in solid $^4$He as the
temperature is decreased, as seen in
Fig.~\ref{mob}. At low temperatures, 
below 200mK, solid $^4$He crystals stiffen considerably. This is 
thought to be due to pinning of dislocation network by $^3$He impurities
and damping of dislocation motion due to phonon collisions. 
Within our assumptions, however, %sull4
we find that as the superfluid fraction increases at lower %sull3
temperatures the dislocation mobility 
decreases resulting in the stiffening of the crystal. Numerical estimates
of the change in shear modulus and the 
$Q$ factor based on this effect model the  
experimental behavior quite well, as can
be seen from the figures. The quantitative agreement might 
 be even better 
when it is taken into account
that bulk viscosities 
of solid $^4$He are likely be larger than the values
for liquid $^4$He used in the mobility estimates. 
We have used the known values of these quantities for the
liquid, as information  for their values in the solid
is lacking. 

Our results  then
show that quenching of dislocation motion due to a
superfluid field could be an important source, even  
possibly the dominant one, of damping for dislocation motion in clean crystals at the %sull3 
low temperature limit.
In this limit (i.e. $T \leq 0.04K$), we find that
the superfluid contribution to dislocation damping is 
likely to be considerably larger 
than that  due to other  
sources of damping (phonons and impurity pinning). 
The superfluid contribution would dominate because, experimentally, 
it is seen that limitation of dislocation motion in conventional
crystals due to Peierls barrier is absent in solid $^4$He. 
The onset temperature of this unusual elastic behavior in
solid $^4$He 
is in the same temperature range as the onset of
superfluid behavior in~\cite{shev,aleinikava} dislocation 
networks.
As noted above, the superfluid contribution to dislocation mobility 
scales with the superfluid fraction i.e. ${\rho_s}/{\rho}$. 
The onset temperature of  the superfluid fraction is reflected, 
in our model, in  the temperature dependence of the shear
modulus effect.
This 
enables us to fit both the magnitude and the temperature 
dependence of the change in shear 
modulus of solid $^4$He crystal. However,
at higher temperatures, the 
factor ${\rho_s}/{\rho}$ becomes smaller 
and eventually approaches zero. In that range, other contributions to 
damping (chiefly 
due to $^3$He impurities and also phonon collisions) will become more important. %sull3 %sull4
Therefore, we believe that the interplay of these effects - superfluid field, 
$^3$He impurities and phonon collisions - 
should be considered in understanding the anomalous softening of the $^4$He crystal. 

%CD added acknowledgement
\begin{acknowledgement} 
This research was supported in part by IUSSTF grant 94-2010.
\end{acknowledgement}


\begin{thebibliography}{99}
\bibitem{note} The acceleration time constant is given by the ratio of mass per unit length  %abd1 added note
to the inverse mobility of the dislocation line. This time constant is very small
$\sim 10^{-13}$ sec.
\end{thebibliography}
\end{document}